\documentclass[12pt]{article}
\usepackage{pic04}
\usepackage{hyperref}
\usepackage{url}
\usepackage{graphicx}

\begin{document}

\title{\bf $B_d$ MIXING AND PROSPECTS FOR $B_s$ MIXING AT D\O }
\author{
Tulika Bose \\
{\em Columbia University, New York, New York, NY 10027, USA} \\
for the D\O\ Collaboration \\}

\maketitle

\baselineskip=14.5pt
\begin{abstract}
Measurement of the $B_s$ oscillation frequency via $B_s$ mixing analyses provides a powerful constraint on the CKM matrix elements. The study of $B_d$ oscillations is an important step towards $B_s$ mixing and a preliminary measurement of $\Delta m_d$ has been made with $\sim$ 250 pb$^{-}$ of data collected with the upgraded Run II D\O\ detector. Different flavor tagging algorithms have been developed and are being optimized for use on a large set of $B_s$ mesons that have been  reconstructed in different semileptonic decay modes.
\end{abstract}

\baselineskip=17pt
\section{Introduction}
First observed in the  $B^0_d - \bar{B^0_d}$ meson system at ARGUS, the neutral B meson transition from the particle to anti-particle state, and vice versa, occurs through a second order weak transition or ``box diagram''. The frequency of the oscillation is proportional to the  small difference in mass between the two eigenstates, $\Delta m$,  and for the $B^0_d - \bar{B^0_d}$ system can be translated into a measurement of the CKM element $|V_{td}|$. $|V_{td}|$ can be used to constrain the unitarity triangle and thereby yield information on the {\it CP} violating phase \cite{abbaneo}. $\Delta m_d$ has been precisely measured (the world average is $\Delta m_d=0.502 \pm 0.007 \hspace{0.1in}ps^{-1}$  \cite{pdg}) but large theoretical uncertainties dominate the extraction of $|V_{td}|$ from $\Delta m_d$. This problem can be reduced if the $B_s^0$ mass difference, $\Delta m_s$, is also measured. $|V_{td}|$  can then be extracted with better precision from the ratio:
\begin{equation}
\frac{\Delta m_s}{\Delta m_d}=\frac{m(B_s^0)}{m(B_d^0)}\xi ^2 {\mid \frac{ V_{ts}}{V_{td}}\mid}^2
\end{equation}
where $\xi$ is estimated from Lattice QCD calculations to be $1.15^{+0.12}_{-0.00}$  \cite{schneider}. The above has motivated many experiments to search for $B_s^0$ oscillations and though a statistically significant signal hasn't been observed yet, a lower limit ($\Delta m_s >14.4 $ ps$^{-1}$ at 95\% C.L.) has been set. Since this current limit indicates that the $B_s^0$ oscillations are at least 30 times faster than the $B_d^0$ oscillations, a $B_s^0$ mixing measurement is experimentally very challenging.

\section{Experimental considerations}
\begin{figure}[htbp]
  \centerline{\hbox{ \hspace{0.2cm}
    \includegraphics[width=5.5cm]{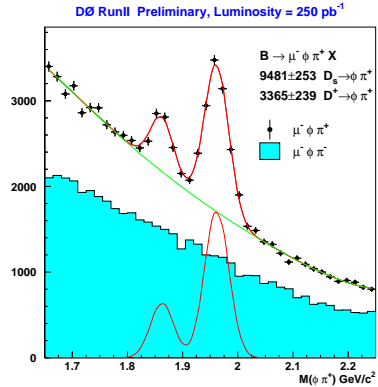}
    }
  }
 \caption{\it $B_s^0\rightarrow D_s^- \mu^+ X ( D_s^- \rightarrow \phi \pi^-)$ yield.
    \label{dsphipi} }
\end{figure}

The D\O\ experiment at the Fermilab Tevatron, a $p\bar{p}$ collider at 1.96 TeV center of mass energy, is well equipped to search for $B_s^0$ oscillations. The large muon acceptance and forward tracking coverage of the D\O\ detector (pseudorapidity coverage of $|\eta|< 2.0$ for the muon, $|\eta| < 1.7$ for the tracking and $|\eta| < 3.0$ for the silicon sub-detectors), along with a robust muon trigger are highly effective in exploiting  the large  $b\bar{b}$ cross-section resulting in some of the largest semileptonic $B_s^0$ yields. Fig.~\ref{dsphipi} shows the yield  for the decay $B_s^0\rightarrow D_s^- \mu^+ X ( D_s^- \rightarrow \phi \pi^-)$ \footnote{Conjugate modes are implied throughout the paper} in $\sim$250 pb$^{-1}$ of data. The decay $B_s^0\rightarrow D_s^- \mu^+ X ( D_s^- \rightarrow K^{*0} K^-)$ has also been reconstructed and is expected to contribute significantly to the total $B_s^0$ yield. Other decays including hadronic $B_s^0$ decays are being studied as well.
%

Other elements essential to a mixing analysis are given by the expression for the average statistical significance:
\begin{equation}
Significance (\Delta m_s, \sigma_t)=\sqrt{\frac{S\epsilon D^2}{2}}\sqrt{\frac{S}{S+B}}e^{-(\Delta m_s \sigma_t)^2/2}
\label{significance}
\end{equation}
where $S$ is the number of signal events, $\epsilon D^2$ is a measure of the effectiveness of a flavor tagger ($\epsilon$ is the tagging efficiency while $D$ is the ``Dilution''), $B$ is the number of background events and $\sigma_t$ is the proper time resolution.

As indicated above, tagging the meson flavor ($B_s^0$ or $\bar{B_s^0}$) at decay time (final-state tagging) and at production time (initial-state tagging) are crucial.  For the semileptonic modes used for mixing studies at D\O\ the final state particles provide the decay-time tag. For initial-state tagging, different techniques are being studied and optimized by doing measurements of $\Delta m_d$. Three tagging algorithms have been developed so far: opposite-side muon tagging, opposite-side jet charge tagging, and same-side (soft-pion) tagging.
 
\begin{figure}[htbp]
  \centerline{\hbox{ \hspace{0.2cm}
    \includegraphics[width=6.0cm]{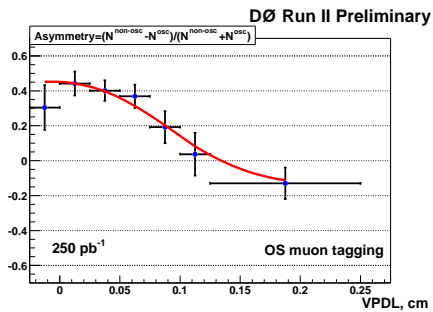}
    }
  }
 \caption{\it The asymmetry between the non-oscillated and oscillated  mesons as a function of the visible proper decay length.
    \label{mixing_slt} }
\end{figure}

The muon tagger relies on identifying the flavor of the other B meson in the event using the sign of the muon it decayed to - a negative muon corresponds to a $b$ quark, and vice-versa.  For the decay used for $B_d^0$ mixing studies - $B_d^0 \rightarrow D^{*-}\mu X$ where $D^{*-} \rightarrow \bar{D^0}\pi^{-}$ and $ \bar{D^0}\rightarrow K^+\pi^-$ - both muons having the same sign would indicate that one B hadron had oscillated while opposite signs would indicate that neither (or both) had oscillated. Fig.~\ref{mixing_slt}  shows  the measured asymmetry between the non-oscillated and oscillated mesons as a function of the visible proper decay length (VPDL). The fit to the asymmetry gives $\Delta m_d = 0.506 \pm 0.055 \pm 0.049$ ps$^{-1}$ with an efficiency of $4.8 \pm 0.2$ \% and $D =  46.0 \pm 4.2$ \% .

The opposite-side jet charge tagging algorithm calculates a $p_T$-weighted average charge of tracks: ${\tt jetQ} \equiv \frac{\sum p_T\,^i q^i} {\sum p_T\,^i}$. A $|{\tt jetQ}| < 0$ value corresponds to a $b$ quark for the B meson on the opposite side, and vice-versa. The same-side pion tagging algorithm makes use of the fact that the charge of the fragmentation pion correlates with the flavor of the reconstructed B meson. A positive pion corresponds to a $\bar{b}$ quark (i.e. $B_d^0$) if the reconstructed $B$ meson is neutral, but to a $b$ quark (i.e. $B^-$) if the reconstructed $B$ meson is charged, and vice versa. Oscillations have been seen using both the opposite-side jet-charge and the same-side pion taggers, and work is ongoing to compute systematics errors and optimize performance. Lastly, since the significance scales as $e^{-(\Delta m_s \sigma_t)^2/2}$, for large values of $\Delta m_s$, a precise measurement of the proper decay time is crucial. Efforts are ongoing in this direction and we hope to have preliminary $\Delta m_s$ results soon.

\section{Conclusions}
 A preliminary measurement of $\Delta m_d$ has been made with the upgraded D\O\ detector. The opposite-side soft muon tagger was used to obtain  $\Delta m_d = 0.506 \pm 0.055 \pm 0.049$ ps$^{-1}$. This measurement, along with the development and optimization of other tagging algorithms, and the reconstruction of different $B_s\rightarrow D_s^{\mp}\mu^{\pm} X$ decay modes, is an important step towards a $B_s$ mixing measurement.

\end{document}